\documentclass{PoS}

\title{The Swift UVOT grism calibration and example spectra}

\ShortTitle{Swift UVOT grism calibration}

\author{Paul Kuin, \speaker{Alice Breeveld}, and Mat Page \\
        Mullard Space Science Laboratory - University College London\\
        E-mail: \email{a.breeveld@ucl.ac.uk}\\
        E-mail: \email{n.kuin@ucl.ac.uk}\\
        E-mail: \email{m.page@ucl.ac.uk}}

\abstract{ 
The calibration of the two UVOT grisms which provide slitless spectroscopy in 
the 170-500 nm (UV grism) and 295-660 nm (visible grism) ranges has been 
completed. The UV grism has a spectral resolution ($\lambda/\Delta\lambda$) of 75 at 
$\lambda$2600~\AA\  for source magnitudes of u=10-16 mag, while the visible grism has a 
spectral resolution of 100 at $\lambda$4000~\AA\  for source magnitudes of b=12-17 mag. 
For brighter spectra, coincidence loss (pile-up) occurs in the photon-counting detector.  
A correction for the coincidence loss in grism spectra has been developed,  and 
limits have been established above which that correction fails. 
After discussing the UVOT grisms and their calibration, 
an illustration is given of the breadth of the UVOT grism spectroscopy. 
 }

\FullConference{Swift: 10 Years of Discovery\\
                 2-5 December 2014\\
                 La Sapienza University, Rome, Italy}

\def \uv{UV\ }

\begin{document}

\section{Introduction}

The {\em Swift} Ultraviolet and Optical Telescope (UVOT) 
includes two grisms in its filter wheel. One was optimised for the 
UV, one for the optical, though the optical grism extends further into 
the UV than is accessible from the ground.  Since the {\em Swift} launch in 
December 2005, grism observations of many targets have been made. 
The easy scheduling of {\em Swift} has made UV spectroscopy of many novae and 
supernovae possible, often within a day of their discovery. 
Spectroscopy of other variables such as Be-WD binary systems 
has also been carried out but is more challenging as these are generally 
fainter \cite{6}. 
Spectroscopy of comets has also been successful in identifying the production rates 
of various molecules and dust from the spectra of their coma \cite{7}. 


\section{The UVOT grisms and their calibration}
\label{cal}

In Table \ref{table1} the relevant parameters of the UVOT grisms have been 
summarized. 
Both grisms are sensitive below the typical atmospheric cutoff at around 3200\AA. 
Although there are only two grisms present in the UVOT filterwheel, each 
grism has two default observing modes, called 'nominal' and 'clocked'.
In the clocked mode part of the aperture is blocked so that for part of the 
image the first orders are free from zeroth order contamination due to field stars.  
The photon-counting detectors provide stable, accurate measurements of the count rate 
over a range of about 5 magnitudes.

\setcounter{table}{0}
\begin{table*}
\begin{center}
   \begin{minipage}{0.65\textwidth}
      \caption{Properties of the UVOT Grisms}
      \label{table1}
   {
\scriptsize
      \begin{tabular}{@{}lccl}
          \hline
                                    &{\bf visible grism } &{\bf \uv grism } \\
          \hline
first order wavelength range       & 2850-6600  \AA                 &1700-5000     \AA \\ 
first order wavelength accuracy (1$\sigma$)    & 22         \AA                 & 9 (18$^a$) \AA \\
spectral resolution                & 100        at 4000~\AA         & 75           at 2600~\AA  \\  
no order overlap (first order)     & 2850-5200  \AA                 & 1700-2740    \AA \\
effective magnitude range          & 12-17      mag                 & 10-16        mag \\  
dispersion (first order)           & 5.9        \AA/pixel at 4200\AA& 3.1          \AA/pixel at 2600\AA \\
zeroth order $b$-magnitude zeropoint & 17.7     mag                 & 19.0        mag\\ 
effective area error nominal mode    & 11 \%                        & 15 \%  \\      
effective area error clocked mode    & 15 \%                        &  9 \%  \\  
         \hline
     \end{tabular}   
     \medskip
     {    
     \begin{tabular}{@{}ll}
$a$\  &UV nominal grism.\\
     \end{tabular} 
     }
   }
   \end{minipage}
\end{center}  
\end{table*}

The grism provides slitless spectroscopy and the spectrum position on the image is 
referenced by using an anchor point. 
The position of the anchor on the image is derived from the sky position of the source 
and the pointing knowledge. 
One can use either the zeroth orders in the grism image, or an image 
in a lenticular filter taken along with the grism exposure to determine the pointing and 
the anchor position. 
The lenticular filter method being preferred, as without it the wavelength errors are currently 
about twice as large. 
The wavelength accuracy given in Table \ref {table1} is based on using the 
lenticular filter. 
The anchor position accuracy is the main source for the wavelength error, with possibly 
a small additional error at the end of the wavelength range as a result of the 
non-linear dispersion sensitivity to the anchor. 

\begin{figure}
   \begin{center}
   \includegraphics[width=0.76\textwidth,angle=0]{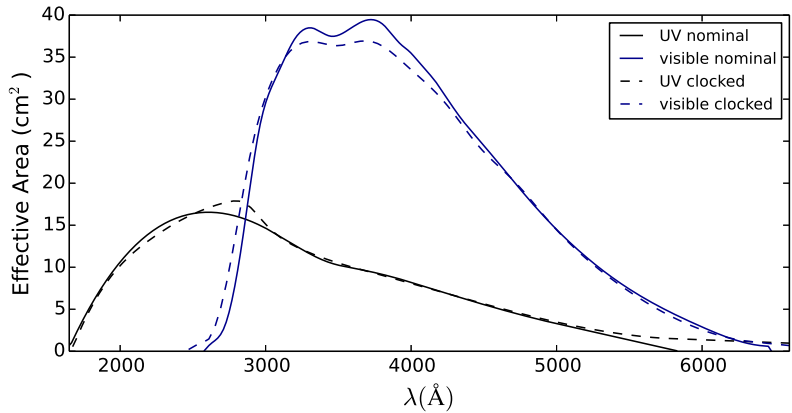}
   \caption{{\small
   The effective areas. The effective areas for the clocked modes are only valid for spectra 
   near the centre of the detector.
   }}
   \label{effarea}
   \end{center}
\end{figure}

The UVOT microchannel plate intensified photon counting detector system is read out  
every 11 ms and thus the image is being built up. 
If in one such 11ms time frame there are multiple photons incident at the same detector 
location, only one is registered. Therefore there are lost counts, and this is known as 
coincidence loss. The process is governed by Poisson statistics, and a correction can 
be made. We have calibrated this correction which affects the 
brightest sources and/or features in the grism spectra.  The correction is made by 
multiplying the observed count rate with a factor which is accurate to within 20\%. 

The grism throughput varies with the wavelength of the photons. 
This sensitivity variation is determined in terms of the grism effective area. 
The effective area was determined for each grism mode after correcting for coincidence loss.  
The nominal grism modes show a nearly constant effective area over the whole detector. 
In the clocked grism modes, due to the clocking covering part of the aperture, 
the effective area varies by position of the spectrum on the detector. 
Fig. \ref{effarea} shows the resulting effective areas for spectra at the 
centre of the detector.  

The calibration of the grisms is described in detail in \cite{3}.
In the next section a sample of the spectra can be found. It should be mentioned that an easy to use spectral 
extraction is now possible by using the {\tt UVOTPY} 
software \cite{2} which was written in the freely available 
Python language. 
Documentation for the grism\footnote{http://www.mssl.ucl.ac.uk/www\_astro/uvot} 
and the {\tt UVOTPY} software\footnote{http://github.com/PaulKuin/uvotpy} are
available on-line.

\section{A display of grism spectra}

\begin{figure}
   \begin{center}
      \includegraphics[width=0.8\textwidth,angle=0.]{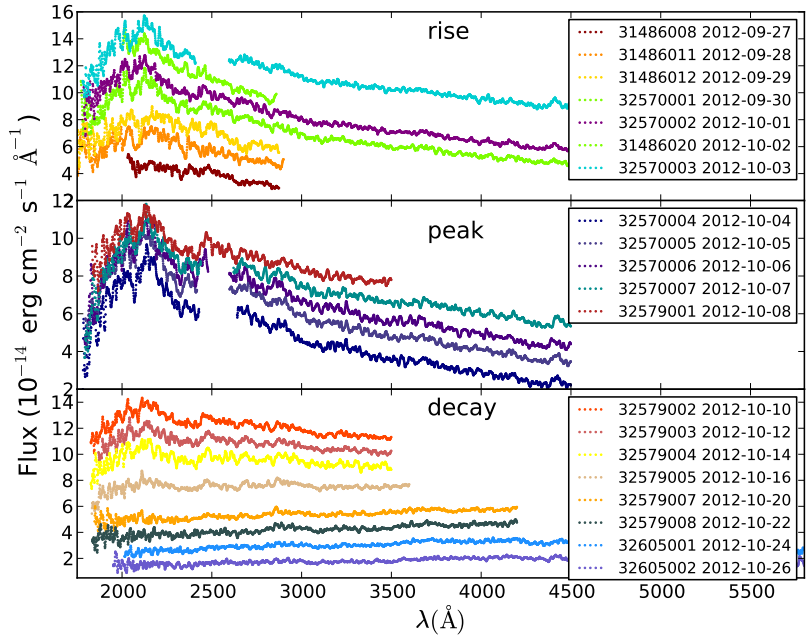}
      \caption{{\small
      The evolution of the spectrum of SN2009ip observed with the UV grism.
      The spectra have not been offset; the brightness variations are real.
      }}
      \label{SN2009ip}
   \end{center}
\end{figure}

\subsection{SN2009ip}

The Type IIP supernovae have substantial UV emission which evolves over time 
when the ejecta cools. A good example is SN2009ip, see Fig. \ref{SN2009ip}, 
\cite{4} which had a failed eruption in 2009 and was then suspected as being a 
SN imposter, but eventually became a well-observed full-blown supernova in 2012. 
The required exposure times became longer when the SN became fainter. The last 
spectra were obtained by summing multiple exposures. Early exposures 
had the spectrum positioned at the centre of the detector. By using an offset 
position, later spectra avoided second order contamination up to much longer 
wavelengths, which can clearly be seen in the spectra. Some gaps were created where 
zeroth orders of field stars contaminated the spectra. 

\begin{figure}
\begin{center}
\includegraphics[width=0.76\textwidth,angle=0.]{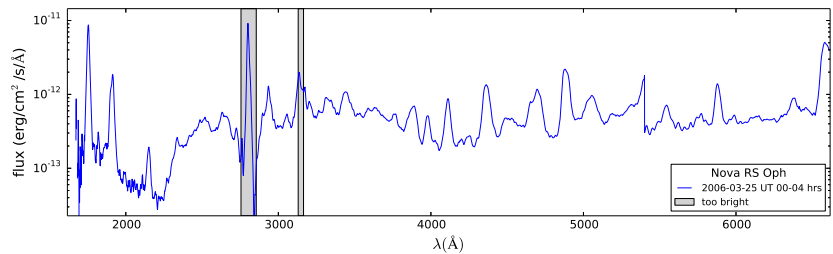}
\caption{{\small
Nova RS Oph 2006. The spectrum is a combination of 3 UV grism spectra 
and 3 visible grism spectra, with omission of the overexposed parts in the visible
grism spectrum below 5400\AA. The feature shortward of 5400\AA\  is due to the second order 
3135 line in the UV grism which was cut off above 5400\AA. 
}}
\label{nova}
\end{center}
\end{figure}

Many supernova spectra have been observed with the UV grism. Their spectra 
are being carefully processed and will be made available through the 
SOUSA archive \cite{1}.

\subsection{Nova RS Oph 2006}

Nova RS Oph was observed with both grisms during the 2006 outburst. Combining the
data from both grisms, a spectrum from 1750 to 6600\AA\  has been obtained as shown 
in Fig. \ref{nova}. 
Note that the Mg II 2800, and (marginally) the 3135\AA\  lines were too bright; 
the count rate there is too large for a coincidence loss correction in both grisms.  
Such a bright feature can also be seen to affect the nearby continuum by depleting the 
observed count rate there. This is related to the spatial extent of the coincidence 
loss. 


\begin{figure}
\begin{center}
\includegraphics[width=0.76\textwidth,angle=0]{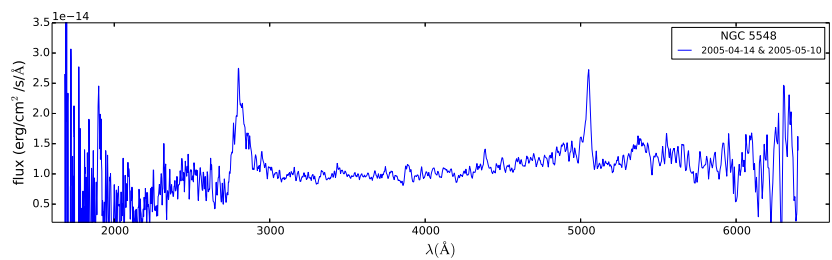}
\caption{{\small 
The spectrum of an Active Galactic Nucleus (NGC 5548)
}}
\label{AGN}
\end{center}
\end{figure}

\subsection{AGN}

The spectrum of NGC~5548 (April 2005) of an active galactic nucleus, 
is shown in Fig. \ref{AGN}.
The spectrum shown is the sum of a UV and visible grism spectrum.  
Notable are the emission lines of CIII], Mg II, and [OIII]. Detailed modeling 
of the NGC 5548 spectrum, in combination with other multi-spectral data, 
can be found in \cite{5}.

\subsection{Wolf-Rayet Stars}

\begin{figure}
\begin{center}
      \includegraphics[width=0.76\textwidth,angle=0]{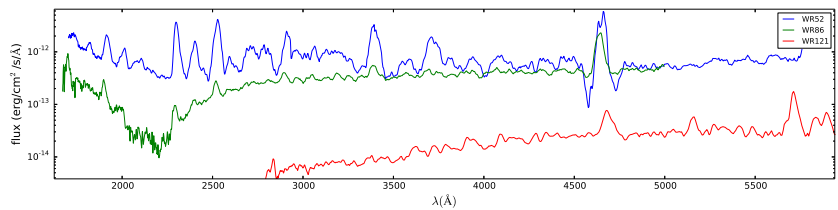}
      \caption{Wolf-Rayet stars WR52, WR86, and WR121.}
      \label{wr}
\end{center}
\end{figure}

Spectra of a number of galactic WR stars have been obtained in a fill-in program. 
The brighter WR~52 and WR~86 were observed earlier by IUE, but we have
obtained a new spectrum of the weaker WR~121, see Fig. \ref{wr}. 
These spectra span approximately the brightness range accessible 
with the UVOT grisms.   

\begin{figure}
   \centering
   \begin{minipage}{0.76\textwidth}
      \centering
      \includegraphics[width=1.0\textwidth,angle=0]{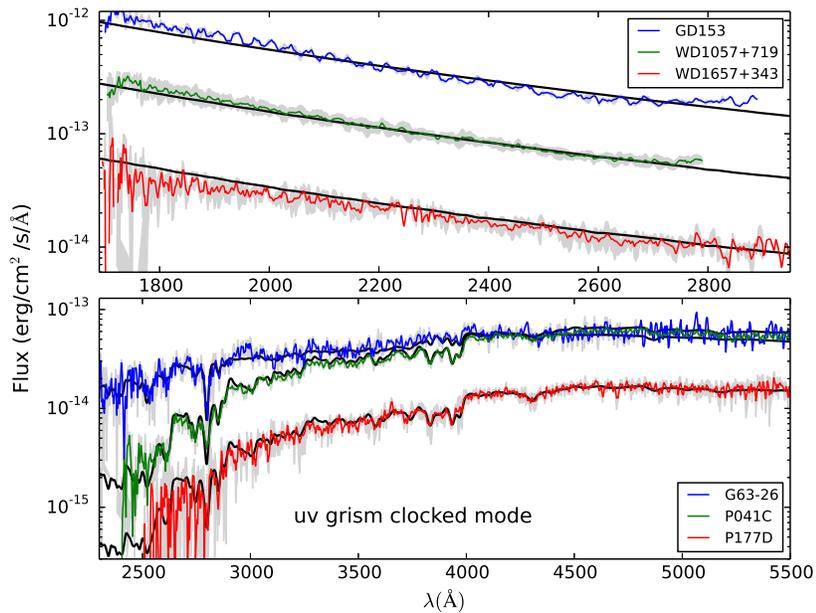}
      \caption{Comparison of the grism spectra to HST standards. 
      The HST spectra are in black; uvot spectra coloured.}
      \label{standards_comp}
   \end{minipage}\hfill
\end{figure}

\subsection{Comparison to HST spectral standards}

For the calibration of the coincidence loss and the effective area, grism spectra
have been compared to HST standards. In Fig. \ref{standards_comp} the HST and Swift
UVOT spectra are plotted together for White Dwarfs and Solar type stars. The 
Solar type stars do not have much UV flux, while the UV dominates in the White Dwarf
spectra. This means that for the Solar type stars the second order contamination 
can be neglected for wavelengths below 4500\AA, while in the White Dwarfs 
the second order can affect the spectrum from 2750\AA\ and above.

\end{document}